\documentclass{article}
\usepackage{amsmath}%mathtime
\usepackage[pdftex]{graphicx}

\newcommand{\ave}[1]{\langle #1\rangle}
\newcommand{\eq}[1]{Eq.\,\ref{#1}}

\newcommand{\integrate}[3]{\int_{#1}^{#2}\mathrm{d}{#3}\,}

\begin{document}

%\onecolumn

%\setcounter{page}{1} %first page number
\title{Quantifying mRNA synthesis and decay rates using small RNAs}

\author
{Vlad Elgart, Tao Jia, Rahul Kulkarni,\\
Department of Physics, Virginia Tech,\\
E-mail: elgart@vt.edu, kulkarni@vt.edu.}

\maketitle

%\pagestyle{headings}

%\markboth{Biophysical Journal: Biophysical Letters}{Biophysical Journal: Biophysical Letters} %for running head

%Abstract environment needs 3 arguments. They are
%1. The abstract
%2. Received date
%3. Address, email

\begin{abstract}
{Regulation of mRNA decay is a critical component of global cellular
adaptation to changing environments. The corresponding changes in mRNA
lifetimes can be coordinated with changes in mRNA transcription rates
to fine-tune gene expression. Current approaches for measuring mRNA
lifetimes can give rise to secondary effects due to transcription
inhibition and require separate experiments to estimate changes in
mRNA transcription rates. Here, we propose an approach for
simultaneous determination of changes in mRNA transcription rate and
lifetime using regulatory small RNAs to control mRNA decay. We analyze
a stochastic model for coupled degradation of mRNAs and sRNAs and
derive exact results connecting RNA lifetimes and transcription rates
to mean abundances. The results obtained show how steady-state
measurements of RNA levels can be used to analyze factors and
processes regulating changes in mRNA transcription and decay.}

\end{abstract}

\vspace*{2.7pt}

%\begin{multicols}{2}
Cellular adaptation to changing conditions is critically dependent on
processes that enable rapid responses to environmental
fluctuations. While considerable research has focused on changes in
transcription, research over the past several years has demonstrated
that control of mRNA decay plays an increasingly important role in
cellular responses \cite{bernstein02,condon07,garneau07}. Correspondingly, there is significant interest in
understanding factors and processes which govern regulation of mRNA
decay.

The traditional approach for measuring mRNA lifetimes involves
quantification of mRNAs remaining at different times following
inhibition of transcription, e.g., by the addition of rifampicin
\cite{condon07}. This procedure requires multiple measurements during
time intervals of the order of the mRNA lifetime, hence high temporal
resolution is required for short-lived mRNAs. More significantly, the
procedure for inhibition of transcription can give rise to secondary
effects which influence mRNA decay \cite{condon07}, hence it is of
interest to consider alternative approaches. In the following, we
outline a novel proposal for quantifying mRNA decay.

A naturally occurring process which regulates turnover for many
bacterial mRNAs involves interactions with regulatory small RNAs
(sRNAs). Experiments have shown that post-transcriptional regulation
of gene expression can occur via binding and subsequent {\em coupled}
degradation of the mRNA and regulatory sRNA which occurs rapidly
\cite{masse03,waters09}. The corresponding coarse-grained model of
gene expression has been analyzed by several groups
\cite{levine07,levine08,mehta08,mitarai07,mitarai09} and shown to be
consistent with experimental observations \cite{levine07}. We will
show that the kinetic scheme for this mode of regulation leads to an
{\it exact} mathematical result relating RNA decay rates. The obtained
result is valid even in the presence of large fluctuations that are
typical for low abundance mRNAs.

\begin{figure}[t!]\vspace*{3pt}
\centering 
\includegraphics{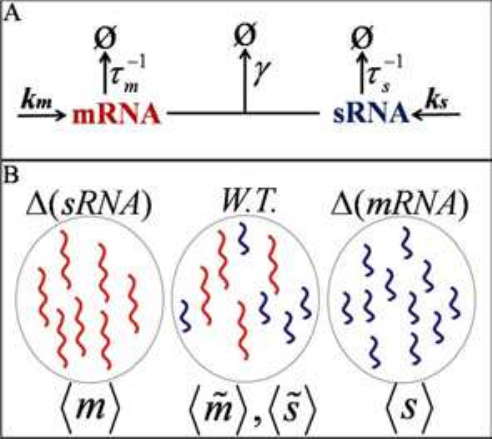}
\vspace*{-5.5pt}
\caption{Panel A: Kinetic scheme for mRNA and sRNA production and
decay including coupled mutual degradation.  Panel B: The proposed
setup involves steady-state measurements for three strains:
$\Delta$(sRNA), WT and $\Delta$(mRNA).  Unregulated steady-state mean
levels of mRNAs and sRNAs along with regulated levels of these
molecules are measured. The measured quantities allow
determination of the average mRNA transcription rate $k_m$ and decay
rate $\tau_m$ relative to the sRNA production rate $k_s$. If $k_s$ is
held fixed, and the conditions are varied, the proposed scheme leads
to simultaneous determination of fold-changes in the rate of
transcription and the rate of mRNA decay. Note that the mRNA/sRNA
interaction parameter can be arbitrary.}
\vspace*{-8pt}
\end{figure}

Our analysis considers a generalized stochastic model for regulation
by sRNAs as illustrated in Fig. 1.  The complexity of processes
leading to transcription suggests that RNA synthesis in many instances
is not adequately modeled as a Poisson process \cite{kaufmann07,raj08}.  
Hence, we model mRNA and sRNA production by {\em arbitrary}
stochastic processes with mean arrival rates $k_{m}$ and $k_{s}$
respectively. Degradation of a mRNA and a sRNA can occur either
independently (with constant probability per unit time $\frac{1}{\tau_{m}}$ 
and $\frac{1}{\tau_{s}}$ respectively) or through the coupled degradation 
process.

The experimental setup in our proposal for quantifying mRNA decay is
as follows. Consider three different strains as shown in Fig. 1B: two
unregulated strains (i.e., with either sRNA or mRNA deleted) and the
wild type (WT) strain. In the WT strain, both mRNA and sRNA are
present and regulate each other.  In steady state, we derive the
following exact relations (Appendix) connecting mRNA/sRNA lifetimes
and transcription rates to the mean abundances
\begin{eqnarray}
	k_{m}~\tau_{m} &=& \ave{m} \nonumber \\
	k_{s}~\tau_{s} &=& \ave{s} \label{s}
\end{eqnarray}
for the unregulated strains, and
\begin{equation}
	\frac{\tau_m}{\tau_s}  = \frac{\ave{m} - \ave{\tilde m}}{\ave{s} - \ave{\tilde s}}. \label{theorem}
\end{equation}

%\doiline
%\end{multicols}

%\twocolumn

Here $\ave{m}$ and $\ave{s}$ are the mean mRNA(sRNA) abundances in
strains lacking the sRNA(mRNA), and $\ave{\tilde m}$ and $\ave{\tilde
s}$ are the mean mRNA and sRNA levels in WT strain where both are
present.

The above relations suggest an alternative approach (Fig. 1B) for
quantifying decay times for mRNAs that either have a naturally
occurring small RNA regulator or for which an antisense RNA regulator
can be designed. Consider an experimental setup expressing the sRNA
from an inducible promoter such that its transcription rate is
primarily controlled by inducer concentration. The mean transcription
rate $k_s$ can, in principle, be determined using single-molecule
methods \cite{yu06}. Now the basic parameters for the coupled system
are $k_m$, $k_s$, $\tau_m$ and $\tau_s$. If $k_s$ is known, then the
values of the other parameters can be determined using experimental 
measurements of
$\ave{m}$, $\ave{s}$, $\ave{\tilde m}$ and $\ave{\tilde s}$ in
combination with equations Eqns. 1-3 above. Alternatively, experiments
can be designed to keep $k_s$ fixed while factors regulating mRNA
decay are changed e.g., by deletion of a protein known to play a role
in mRNA decay. The above equations can be used to simultaneously
determine the corresponding fold-changes of the mRNA/sRNA lifetimes
and the mean mRNA transcription rate.

The proposed approach can be used to address several important
questions of current interest, some of which are highlighted in the
following. By targeted mutagenesis of specific mRNA sequence elements,
the induced fold-change in mRNA lifetime, as well as the corresponding
change in the transcription rate $k_m$, can be determined using the
same experimental setup. This is an important feature, given that
recent experiments have observed coordination between changes in
transcription and changes in mRNA degradation \cite{shalem08}.
Quantifying the change in mRNA lifetimes induced by mutations to
different components of cellular degradation pathways can address such
issues as the role of polyadenylation in mRNA decay \cite{joanny07}.
It would also be of interest to design high-throughput experiments for
different mRNAs which are regulated by corresponding antisense RNAs,
all of which are expressed from identical inducible promoters and thus
have the same $k_s$. The proposed procedure can then be used for
genome-wide determination of relative transcription rates and
lifetimes of mRNAs. These effective parameters, in turn, serve as
critical inputs to systems-level models of cellular processes
\cite{ronen02}.

In summary, we have proved an exact relation for a nonlinear
stochastic model of cellular post-transcriptional regulation. The
derived results suggest a novel procedure for simultaneous
determination of mRNA production rates and mRNA lifetimes. While the
focus was on bacterial mRNAs, the procedure outlined can also be
applied to higher organisms and used to systematically explore the
sequence determinants and processes involved in regulation of mRNA
decay.

\section*{Appendix}

The proposed experimental setup involves measurements of mRNA/sRNA
abundances for three strains, the sRNA deleted $\Delta$(sRNA) strain,
mRNA deleted $\Delta$(mRNA) strain, and the wild type (WT) strain
(both species are present and regulate each other.)
The regulation model (with constant creation rates for mRNA and sRNA)
has been analyzed by several groups
\cite{levine07,mehta08,mitarai09} and is summarized in the reaction
scheme described in Fig.1.

Recent experiments which provide evidence for transcriptional bursting
\cite {raj08} point towards the need to go beyond the Poisson process
as a model for RNA synthesis. Accordingly, we take both mRNA and sRNA
creation events as {\em arbitrary} stochastic processes. Degradation
of RNA is assumed to be a Poisson process with rate $\tau_x^{-1}$,
where $\tau_x$ is the mean lifetime of $x$RNA , $x=\{m,s\}$. In the WT
strain, mRNA and sRNA undergo a coupled degradation process. We assume
that this process is symmetric with respect to the number of mRNA and
sRNA molecules involved, e.g., $M+S\rightarrow\emptyset$.

Let us choose a particular realization of the system evolution during
time interval $t=[0,T]$. For large values of $T$, we derive
\begin{equation}
	x(T)-x(0) = C_x(T) - Y(T) - \tau^{-1}_x\integrate{0}{T}{t}x(t),\label{diff}
\end{equation}
where $x(t)$ is the number of molecules of the species $x=\{m,s\}$ at
the time $t$.

In \eq{diff}, $C_x(t)$ is the total number of molecules of the species
$x$ created during system evolution until time $T$, and $Y(T)$ is the
total number of molecules of either species that is {\em mutually}
degraded within the time interval $[0,T]$.  Finally, using the law of large 
numbers, the number of molecules degraded naturally in $[0,T]$
is given by the last term in the \eq{diff}.

Dividing both sides of \eq{diff} by $T$ and taking a limit
$T\rightarrow \infty$ we obtain
\begin{equation}
	\lim_{T\rightarrow \infty} \frac{x(T)-x(0)}{T} = k_x -
	\tau^{-1}_x\ave{\tilde x} - \lim_{T\rightarrow \infty}
	\frac{Y(T)}{T},\label{diff2}
\end{equation}
where $\ave{\tilde x}$ is average number of molecules in the
system. We also defined $k_x$ as a mean arrival rate of the species
$x=\{m,s\}$. The limit on the left hand side of
\eq{diff2} vanishes in the case of finite degradation rates
$\tau^{-1}_x$ (number of molecules at any time is finite.) Note that
$Y(T)$ is an extensive quantity (it is monotonic increasing function
of $T$) and therefore, the limit on the right hand side of \eq{diff2}
is finite.

Hence, we derive
%\begin{equation}
%	0 = k_m - \tau^{-1}_m\ave{\tilde m} - \lim_{T\rightarrow
%	\infty} \frac{Y(T)}{T},\\ 0 = k_s - \tau^{-1}_s\ave{\tilde s}
%	- \lim_{T\rightarrow \infty} \frac{Y(T)}{T},
%\end{equation}
\begin{eqnarray}
        k_m - \tau^{-1}_m\ave{\tilde m} - \lim_{T\rightarrow
	\infty} \frac{Y(T)}{T} &=& 0, \nonumber \\ 
        k_s - \tau^{-1}_s\ave{\tilde s}
	- \lim_{T\rightarrow \infty} \frac{Y(T)}{T} &=& 0,
\end{eqnarray}

which immediately yields the following expression
\begin{equation}
	k_m - \tau^{-1}_m\ave{\tilde m} = k_s - \tau^{-1}_s\ave{\tilde s}.\label{regulated}
\end{equation}

In the unregulated case $Y(T) \equiv 0$ for any $T$, since one of the
RNA species is deleted and there is no coupled degradation. In this
situation one gets
\begin{equation}
	0 = k_m - \tau^{-1}_m\ave{m},\\
	0 = k_s - \tau^{-1}_s\ave{s},
\end{equation}
where $\ave{x},\, x=\{m,s\}$ are the average number of molecules
during unregulated system evolution. Combining the set of equations
above with \eq{regulated}, we derive the results in the main text Eqns
1-3. We note that the derived results are valid even if the binding of
mRNA and sRNA is taken to be reversible and the lifetime of the
mRNA-sRNA complex is finite. Finally, the time average can be
replaced by the ensemble average in the steady state.

We have validated the derived results using stochastic simulations
based on the Gillespie algorithm \cite{gillespie77}.  Production of
RNA molecules was taken to occur in transcriptional bursts \cite{raj08}
i.e. each burst corresponds to the arrival of a random number of RNAs
drawn from a geometric distribution, conditional on the production of
at least 1 RNA molecule \cite{ingram08}.  The waiting-time between
bursts was a random variable drawn from exponential or Gamma
distributions.  As expected, the results from the
simulations were in excellent agreement with the derived analytical
results.

\bibliographystyle{plain}

\bibliography{mrna_decay}

%\section*{SUPPLEMENTARY MATERIAL}

%An online supplement to this article can be found by visiting BJ Online at http://www.biophysj.org.\vspace*{6pt}

\end{document}